\documentclass[prb,preprint,amssymb,showpacs]{revtex4}
\usepackage{bm}
\usepackage{graphicx}
\usepackage{dcolumn}
\newcommand{\dm}{{\rm d}}
\newcommand{\im}{{\rm i}}

\begin{document}

\title{Electronic friction for a slow impurity in an inhomogeneous
metallic system}

\date{\today}

\author{A. Salin\footnote{Retired from Universit\'e de Bordeaux I}}
\affiliation{12, rue Jules Testaud, 33700 M\'erignac, France}
\email{asalin@mailaps.org}

\begin{abstract}
The stopping power of a metal for a slow structureless impurity varies linearly
with projectile velocity. We show that the coefficient of this linear behavior
(friction coefficient) is determined exactly in a static ensemble Kohn-Sham
scheme, by extension of the Shifted Fermi Surface procedure originally derived
for an homogeneous jellium. We prove that the friction coefficient is determined
only by local characteristics of the system. The error incurred when adding a
spurious non-local contribution is illustrated for a simple 1D model.
\end{abstract}
\pacs{82.65.+r, 34.35.+a, 68.49.-h, 79.20.Rf}
\maketitle

\section{Introduction}
The stopping power of a free (homogeneous) electron gas (FEG) for a
structureless atom or ion varies linearly with projectile velocity in the limit
of low velocities (see, e.g., Ref.~\onlinecite {EFR,pu91,ena89,saez99}).
This result has been proved at the level of both linear and non-linear theories
of stopping. The coefficient of this linear behavior is referred to as the
{\em friction coefficient} and the associated process as {\em electronic
friction}. Such a behavior has been observed in experiments for ion stopping
in metals that are expected to mimic a FEG in this respect. Examples are for H,
D and He ions in Al\cite{pri11}, H and He ions in Au and Cu\cite{mar09} or H
and He ions in Al, Zn and Au\cite{mar96}. It has also been observed in grazing
collisions of ions on surfaces, like H and He ions on the Al(111)
surface\cite{win02}.

One important field in which electronic friction might play an important role,
is that of surface chemistry, as a candidate for energy dissipation by
adsorbates. It is crucial to assess its importance for such processes as atomic
and molecular adsorption, diffusion on surfaces, range of hot atoms following
molecular dissociation at surfaces, etc. Then, the adsorbate is immersed into a
strongly non homogeneous electronic density which raises questions as to the
relevance of values of friction coefficients obtained from calculations based
on the FEG model. It is, therefore, desirable to develop methodologies that can
make use of present day band structure codes to evaluate the friction
coefficient. Such a task has been endeavored by Trail {\em et
al.}\cite{trail1}, based on the formulation proposed by Hellsing and
Persson\cite{hp84}. Results have been obtained for H and D on
Cu(111)\cite{trail2,trail3} or for H$_2$ on Cu(111) and N$_2$ on
Ru(0001)\cite{luntz1}. However, some questions are raised by the latter
results. Firstly, the H/Cu results are divergent at a certain value of the
atom-surface distance. Secondly, the friction coefficient for N$_2$/Ru may reach
values an order of magnitude larger than usually obtained from the FEG model.

In the present work, we reexamine the formulation of the friction coefficient
determination from first principles in order to ensure that we get a consistent
first order approximation in the projectile velocity. Our approach is based on
Density Functional Theory (DFT) and extends the methodology developed for
the homogeneous FEG to the inhomogeneous case.

Atomic units are used throughout.
\section{Adiabatic theory}

We consider a medium composed of electrons moving in a periodic array of fixed
nuclei. We are interested in the projectile energy loss due to the electronic
state perturbation  by the projectile motion, the role of the projectile
frozen-lattice interaction (i.e., with the lattice nuclei) being trivial. The
``external'' potential, in which
the electrons move, is composed of two parts: one, $V_{\rm lat}$, due to the
fixed nuclei and the other, $V_{eP}(t)$, to the moving, structureless,
projectile. As the projectile position with respect to
the lattice, $\bm{R}_P$, is time-dependent, the friction coefficient is a
function of $\bm{R}_P$. We suppose that the static problem, with the projectile
at rest, has already been solved and restrict our study to the limit when the
projectile velocity goes to zero, since the friction coefficient is associated with the first order approximation to the stopping in its power series expansion as a function of the projectile velocity. In principle, the density of
the system, as a function of time, may be determined through the time-dependent
Kohn-Sham (KS) scheme of the Time Dependent Density Functional Theory
\cite{runge} (TDDFT, for a general presentation, see e.g.,
Ref.~\onlinecite{casida}).
However, in the low velocity limit, it is not necessary to resort to TDDFT as we
show below.

Our strategy is as follows. In the present section we study the behavior of the system in the adiabatic limit. In the next section we show that the non-adiabatic corrections do not contribute to the friction coefficient.

Let us start with the projectile at rest in the lattice frame without
interaction with the electrons. We could first solve the static problem by
turning on the projectile-electron interaction and, subsequently, set the slow
projectile motion. However, in the adiabatic limit, we may as well first set the projectile
motion, without interaction with the electrons, and determine the transformation
from the lattice to the projectile frame. We then turn on the
projectile-electron interaction adiabatically. The latter procedure is justified since in the adiabatic limit the final state is independent of the evolution path. 
\subsection{Transformation to the projectile frame}
We first consider the projectile moving without interaction with the electrons.
We assume that the projectile velocity, $\bm{v}$, is constant so that the
transformation from the lattice to the projectile frame is Galilean. Let
$\bm{r}_i$ be the position vector of electron $i$ with respect to an origin
fixed in the lattice frame and $\bm{r}_i^P$ w.r.t.\ the projectile. The
N-electron wave-function $\Psi_T(\bm{r}_1,...,\bm{r}_N)$ becomes in the
projectile frame (discarding in this section, for short, the energy phases that
play here no role):
\begin{equation}
\Psi_P(\bm{r}_1^P,...,\bm{r}_N^P;\bm{R}_P(t))=
e^{-\im\sum_{j=1}^{N}\bm{v}\cdot\bm{r}_j^P}\,\Psi_T(\bm{r}_1,...,
\bm{r}_N)
\label{eq0}\end{equation}
with $\bm{r}_i=\bm{r}_i^P+\bm{R}_P(t)$. The density in the target frame, and in
the absence of projectile-target interaction, is exactly known through a static
Kohn-Sham scheme. This static Kohn-Sham scheme is transformed into the
projectile frame by a similar Galilean transformation involving the
multiplication of each KS orbital in the lattice frame, $\varphi_i^T$, by
$\exp\{-\im\bm{v}\cdot\bm{r}^P\}$. As the KS orbitals are Bloch functions:
\begin{equation}
\varphi^P_{\bm{k},i}(\bm{r}^P;\bm{R}_P(t)) = e^{-\im\bm{v}\cdot\bm{r}^P}
\varphi^T_{\bm{k},i}(\bm{r}) 
= e^{\im\bm{v}\cdot\bm{R}_P } e^{\im(\bm{k}-\bm{v})\cdot\bm{r}}\,u_{\bm{k},i}(\bm{r})
\label{eq1}\end{equation}
where $u_{\bm{k},i}$ is a periodic function\footnote{To conserve periodicity,
an array of projectiles is required, as always done in band structure
calculations. In spite of this, we always refer below, for simplicity, to {\em
the} projectile.}. In the absence of a gap close to the Fermi surface:
\begin{eqnarray}
e^{-\im\bm{v}\cdot\bm{r}}\,\varphi^T_{\bm{k},i}(\bm{r}) 
&=& \varphi^T_{\bm{k}-\bm{v},i}(\bm{r}) + e^{\im(\bm{k}-\bm{v})\cdot\bm{r}}\,
\bm{v}\cdot\bm{\nabla}_{\bm{k}}\,u_{\bm{k},i}(\bm{r})
 + 0(v^2) \nonumber \\
&=&  \varphi^T_{\bm{k}-\bm{v},i}(\bm{r}) + 0(v)
\label{eq2}\end{eqnarray}
 When substituting $\varphi^T_{\bm{k}-\bm{v},i}$ for $\varphi^P_{\bm{k},i}$ (disregarding the common time
dependent phase factor $\exp\{\im\bm{v}\cdot\bm{R}_P\}$), we introduce a shifted KS scheme in the target frame, i.e.\ a
Fermi distribution shifted by $-\bm{v}$ (SFS). It can be shown that this shifted KS scheme is correct to first order in the velocity. Firstly,
the total number of states within the shifted Fermi surface is conserved up to first order in
$v$. Indeed, for vanishingly small
 $v$, and in the absence of a gap in the vicinity of the Fermi surface:
\begin{equation}
 \int \dm\bm{k}\, \rho(\bm{k}-\bm{v}) 
 = \int \dm\bm{k}\, \rho(\bm{k})
+ \int \dm\hat{\bm{k}}_F (-\bm{v}\cdot\hat{\bm{k}}_F)\, \rho(\bm{k}_F)
+ 0(v^2)
\label{eq4}\end{equation}
where $\rho$ is the density of levels and the volume integrals are over the
volume bounded by the unshifted Fermi surface. Since the Fermi surface is
symmetric with respect to the origin of the Brillouin zone, the quantity
$(-\bm{v}\cdot\hat{\bm{k}}_F) \,\dm\hat{\bm{k}}_F$ is exactly compensated by the
same term for $\bm{k}=-\bm{k}_F$. Secondly, the density is also exact to first order in $v$.
\begin{eqnarray}
n^P(\bm{r})&=& \int_{\rm FS} \dm \bm{k}\, \left| \varphi^T_{\bm{k}-\bm{v},i}(\bm{r}) + e^{\im(\bm{k}-\bm{v})\cdot\bm{r}}\,
\bm{v}\cdot\bm{\nabla}_{\bm{k}}\,u_{\bm{k},i}(\bm{r}) \right|^2
+ 0(v^2) \nonumber \\
&=& \int_{\rm FS} \dm \bm{k}\, \left\{ \left|\varphi^T_{\bm{k}-\bm{v},i}(\bm{r})\right|^2
+2\Re \left[ u^*_{\bm{k},i}(\bm{r})\bm{v}\cdot\bm{\nabla}_{\bm{k}}\,u_{\bm{k},i}(\bm{r})
\right] \right\} + 0(v^2) 
\label{eq4a}\end{eqnarray}
The symmetry with respect to the origin of the Brillouin zone implies that $u^*_{\bm{k},i} = u_{-\bm{k},i}$ so that the second term in the integral gives a zero contribution when integrated over $\bm{k}$, which yields:
\begin{equation}
n(\bm{r})= \int_{\rm SFS} \dm \bm{k}\, \left|\varphi^T_{\bm{k},i}(\bm{r})\right|^2
 + 0(v^2)
\label{eq4b}\end{equation}
In other terms, the SFS prescription gives the correct density up to first order in $v$ which means that the corresponding KS scheme is exact to first order in the velocity. We may re-write the previous expression for short, as:
\begin{equation}
n^P(\bm{r}^P;\bm{R}_P(t)) = n^P_{KS}(\bm{r}^P;\bm{R}_P(t))
= \sum_i F_i \left|\varphi_i^T(\bm{r})\right|^2  + 0(v^2)
\label{eq3}\end{equation}
where $F_i$ corresponds to the Fermi distribution shifted by $-\bm{v}$.

As expression (\ref{eq3}) shows, the time dependent KS solution in the
 projectile frame can be formulated in terms of an ensemble KS procedure in the
lattice frame. When $v\ll k_F$ and for a given orientation of $\bm{k}$, the
occupation of orbital $\varphi_{\bm{k}}$ is 1 for $k\le k_m$ and zero for
$k>k_m$, where $\bm{k}_m=k_F\hat{\bm k}-\bm{v}$. The Fermi surface being a
surface of constant energy, this entails that the population decreases as the
energy increases as required for the validity of the Hohenberg-Kohn theorem for
ensembles \cite{dreizler}. Then, the latter theorem proves, in the
non-degenerate case, that the exact state of the system in the projectile frame
is also described by an ensemble in the lattice frame, univocally defined from
the K-S ensemble since the latter determines both the ensemble and the density.

The system we are considering here involves degeneracies, so its ensemble is not
 determined univocally by the density. However, the energy remains a functional
of the density and any quantity that can be expressed in terms of the density is
also univocally defined. These properties are the only ones required for the
validity of the following discussion.

As a conclusion, the Galilean transformation from the target to the projectile
 frame of the exact N-electron state yields, to first order in the velocity, an
ensemble of states which are defined in the (static) target frame. In other
terms, we have transformed the time-dependent problem in the projectile frame
into a static one in the target frame, at the expense of using an ensemble defined by the Shifted
Fermi Surface (SFS).
\subsection{Projectile-target interaction in the adiabatic limit \label{projad}}
\subsubsection{Adiabatic evolution \label{sec:adiab}}
Let $\Psi_j(t_i)$ be the exact solution of a time dependent problem when the
system is in the eigenstate $j$ of the Hamiltonian at the initial time $t=t_i$.
\begin{equation}
H(t_i)\Psi_j(t_i) = E_j(t_i)\Psi_j(t_i)
\label{eq6}\end{equation}
If the evolution is adiabatic,
the system remains in the eigenstate $j$ of $H(t)$ at any time, i.e., it evolves
in such a way that the wave function is at any time the solution of a static
problem. Evolution from two different initial states cannot lead to the same
state at time $t$: adiabatic evolution is reversible. For simplicity, the
derivation below assumes that the eigenenergies are not degenerate\footnote{If
degeneracies occur, as is obviously the case in metallic systems, the evolution
of the subspaces associated with a given energy follows the same trend as in the
non-degenerate case (see, e.g., Ref.~\onlinecite{messiah}, section XVII-II). The
relation between degenerate substates at different times must be done by
continuity. This is trivial if the various substates differ in symmetry, i.e.,
one or more observables, commuting with the hamiltonian, allow to distinguish
the various substates. Even if the associated eigenvalues are time dependent,
this allows to prescribe univocally the adiabatic correspondence between various
times. All KS calculations for an impurity in a solid are periodic (periodic
array of impurities, supercell for surfaces, etc.). Then, the KS eigenstates are
Bloch functions characterized by a $\bm k$ value, which allows to establish the
required correspondence in time.}. Accordingly, the function $\Psi_j(t)$ is
determined at any time $t$ (save for a time-dependent phase factor) by the
energy $E_j(t)$, i.e., the evolution of the system state is determined by the
knowledge of $\dm E_j(t)/\dm t$. This one to one correspondence between energy
and state is central to our derivation below.

The same correspondence prevails for the evolution of an ensemble, which may be
defined by the density operator:
\begin{equation}
\hat{\rho}(t) = \sum_j\, p_j\, |\Psi_j(t)\rangle\langle\Psi_j(t)|
\label{eq6a} \end{equation}
where the weights $p_j$ are constant in time. The energy is given by:
\begin{equation}
E(t) = {\rm Tr} \{\hat{\rho}(t)\,H(t)\}= \sum_j\,p_j\, E_j(t)
\label{eq6b}\end{equation}
Again, the system is at all times characterized by quantities that are the
solution of a static problem. Furthermore, when the $p_j$'s
satisfy the same condition as that for the validity of the HK theorem for ensemble, the
ensemble is determined at any time by its energy.

\subsubsection{Application to the present problem}
In the present problem, the adiabatic evolution is associated with that of the projectile-lattice relative position, i.e.\ with the variable $\bm{R}_P$. What remains to be derived is the KS scheme associated with this adiabatic evolution. To this end, we
switch on adiabatically the projectile-target interaction by writing the
external potential as:
\begin{equation}
v_{\rm ext}(\bm{r},t)= \lambda(t)\,V_{eP}[\bm{r},\bm{R}_P(t)] + V_{\rm lat}
\label{eq5}\end{equation}
The function $\lambda(t)$  is an arbitrary function that varies from 0 to 1 in
the time interval $[t_i,t_f]$. Saying otherwise, we introduce through the
function $\lambda(t)$ a fictitious evolution starting from an initial state
where the projectile and electrons do not interact. Although we could keep $\bm{R}_P$ fixed in this process, we conserve in (\ref{eq5}) its time-dependence. 
As can be seen below, the conclusions of the present section are independent of the presence or absence of the latter time-dependence.
We assume that it is always possible to choose
the arbitrary function $\lambda(t)$ so that $\dm\lambda/\dm t$ is small enough,
at any time, to ensure adiabaticity. The adiabatic approximation
corresponds to a zero order approximation in $\dm\lambda/\dm t$
(see,e.g.,\cite{messiah}).

Our aim is to determine the KS orbitals, at each time, when the system evolves
adiabatically. To this end, we calculate the energy variation in the interval
$\dm t$:
\begin{eqnarray}
{\dm E(t)\over \dm t} &=& {\rm Tr} \{\hat{\rho}(t)\,{\dm H\over \dm t}\}
\hspace{5cm}\nonumber \\
&=&  \int \dm\bm{r}\, n(\bm{r},t)\,\left[ {\dm\lambda\over \dm t}\,V_{eP}
\right.  +\lambda(t)\,\bm{v}\cdot\bm{\nabla}_{\bm{R}_P}V_{eP}\Big]
\label{eq7} \end{eqnarray}
where $n$ is the exact density at $t$. According to the discussion of section
\ref{sec:adiab}, the KS orbitals must be such that they give the exact energy
variation (\ref{eq7}) in the limit of vanishing $\dm \lambda/\dm t$.
We suppose that, at a given time $t^0$, we have determined the KS orbitals for
the ensemble defined by the
distribution $G_i$ and study the time evolution in the interval $[t^0,t^0+\dm
t]$. Consider the orbitals $\varphi_i^a(\bm{r},t)$ evolving adiabatically from
the KS orbitals at $t^0$ for the same ensemble, i.e., assuming that $G_i$ is
independent of time:
\begin{equation}
n^a(\bm{r},t) = \sum_i G_i \left|\varphi_i^a(\bm{r},t)\right|^2
\label{eq8}\end{equation}
where the orbitals $\varphi_i^a$ are the solution of a static problem
\begin{equation}
H_{KS}\, \varphi_i^a(\bm{r},t)
 =  \varepsilon_i^a(t)\,  \varphi_i^a(\bm{r},t)
\label{eq9} \end{equation}
with
\begin{equation}
H_{KS} = -{1\over 2} \nabla^2+ \lambda(t)\,V_{eP}(\bm{r};\bm{R}_P)
+v_n(\bm{r};[n^a(\bm{r},t)])
\label{eq9a} \end{equation}
and $v_n$ includes the Hartree potential, the lattice potential and the exchange
correlation potential of the static problem for the KS ensemble defined by the
occupations $G_i$. The question is now: what is the relation between these {\em
adiabatic KS orbitals} and the KS orbitals associated with the exact adiabatic
evolution of the system? Introduce the energy functional:
\begin{eqnarray}
E_v[n] &=& \sum_i G_i\,\varepsilon_i(t;[n]) - \int \dm\bm{r}\, n(\bm{r},t)\,
v_{\rm xc}(\bm{r};[n(\bm{r},t)]) \nonumber \\
& & { } - {1\over 2}\int\int \dm\bm{r}\,\dm\bm{r}'\,{n(\bm{r},t)\,
n(\bm{r}',t)\over |\bm{r}-\bm{r}'|} + E_{\rm xc}[n]
\label{eq10}\end{eqnarray}
The functional $E_v[n]$ gives the exact energy of the static ensemble with
density $n$, when the $\varepsilon_i$'s, the exchange-correlation potential
$v_{xc}$ and the energy $E_{xc}$ are those of the static KS ensemble defined by
the occupations $G_i$. At $t^0$, $n=n^a=n^0$ and $E_v[n^0] = E(t^0)$ where $n^0$
and $E(t^0)$ are the
exact density and energy. The question is: what is the relation between
$E_v[n^a]$ and $E(t)$ at $t=t^0+\dm t$? We split the variation of $E_v[n^a]$
over the interval $\dm t$ into two terms. The first one corresponds to the
variation of $n^a$:
\begin{equation}
\left.{\dm E_v[n^a]\over \dm t}\right|_{\lambda V_{eP}} =
\int \dm\bm{r}\, \left. {\dm E_v[n^a]\over \dm n^a}
\right|_{n^a=n^0(\bm{r},t^0)}\, {\dm n^a(\bm{r},t)\over \dm t}
\label{eq11}\end{equation}
This term is identically zero because at $t=t^0$ the energy functional is stationary
around the exact density $n^0(\bm{r},t^0)$. The second term corresponds to the
explicit dependence on $\lambda(t)\,V_{eP}$ in (\ref{eq10}) for a fixed $n^a$:
\begin{eqnarray}
\left.{\dm E_v[n^a]\over \dm t}\right|_{n^a} &=&
\sum_i G_i {\dm\varepsilon_i^a(t) \over \dm t} \nonumber \\
&=&  \int \dm\bm{r}\, n^0(\bm{r},t^0)\,\left[ {\dm\lambda\over \dm t}\,V_{eP}
+\lambda(t)\,\bm{v}\cdot\bm{\nabla}_{\bm{R}_P}V_{eP}\right]
\label{eq14}\end{eqnarray}
which is identical to (\ref{eq7}). As a conclusion, at $t^0+dt$ the
$\varphi^a$ orbitals are still the KS orbitals for the exact adiabatic evolution
when the ensemble is defined by the {\em same} time independent distribution
$G_i$. This result may be applied to the evolution over the whole interval
$[t_i,t_f]$. Now, for $t=t_i$, i.e., in the absence of projectile target
interaction ($\lambda(t)=0$), the distribution $G_i$ is exactly given by the SFS
prescription, $G_i=F_i, \forall i$. This means that the solution of the problem
at $t=t_f$ ($\lambda(t)=1$) is also given by the SFS prescription\footnote{This
 result is compatible with a deformation of the Fermi surface in the evolution.
The change in $\bm k$ for a given state, in the adiabatic evolution, is
determined by continuity and
the inversion of (\ref{eq1}) and (\ref{eq2}) ensures the correspondence, to
first order in $v$, at any $t$ between the shifted and a corresponding
``unshifted'' surface. Note that the latter does not correspond to the
solution for the
projectile at rest since the KS equations have been solved for a shifted Fermi
surface. This introduces a difference of order $v$ in the potential which is the
origin of friction. Obviously, the SFS solution is not merely the Galilean
transform of the solution with the projectile at rest.}.

We are led to the following conclusion: the density, in the adiabatic limit, can
be determined, at each position of the projectile, using a {\em static} ensemble
KS scheme in which {\em both} the projectile and lattice potential are fixed.
The fact that the energy is stationary with respect to an arbitrary variation of
the density around its exact value (see Eq.~\ref{eq11}) plays a central role in
reaching this conclusion.
The SFS
Kohn-Sham scheme is an exact scheme for the determination of the system density
for an impurity traveling through an inhomogeneous medium under adiabatic
conditions.
\section{Friction coefficient}

The force acting on the projectile is:
\begin{equation}
\bm{\nabla}_{\bm{R}_P} E(\bm{R}_P) =-\int d\bm{r}\,
n(\bm{r};\bm{R}_P)\, \bm{\nabla}_{\bm{R}_P}V_{eP}(\bm{r};\bm{R}_P)
\label{f:eq3} \end{equation}
which involves only the derivative of $V_{eP}$ since, as for (\ref{eq11}), the
derivative of the energy functional with respect to density is zero.

When the density in (\ref{f:eq3}) is the static density
$n_0(\bm{r};\bm{R}_P)$, the force is the static force, i.e.\ the limit when the projectile velocity $\bm v$ goes to zero. When the
projectile moves with a small but finite velocity $\bm v$, the associated density is
$n_{\bm v}(\bm{r};\bm{R}_P)$ and a dissipative process takes place which
corresponds to friction. The friction coefficient for motion along the direction
$\hat{\bm v}$ is defined by:
\begin{equation}
{\cal F}_{\hat{\bm v}} = \lim_{v\rightarrow 0} {1 \over v}
\left\{ \int d\bm{r}\, \Delta n(\bm{r};\bm{R}_P)\, \hat{\bm
v}\cdot\bm{\nabla}_{\bm{R}_P}V_{eP}(\bm{r};\bm{R}_P) \right\}
\label{f:eq4} \end{equation}
where
\begin{equation}
\Delta n(\bm{r};\bm{R}_P)\ = n_{\bm v}(\bm{r};\bm{R}_P) -
n_0(\bm{r};\bm{R}_P)
\label{f:eq4a} \end{equation}
%
\subsection{Non-adiabatic corrections\label{nonadiab}}
When $n_{\bm v}$ is calculated in the adiabatic SFS approximation of section \ref{projad}, the dependence of $n_{\bm v}$ on $\bm v$ arises entirely from the SFS. However, we have to check whether non-adiabatic corrections can be neglected when $n_{\bm v}$ is calculated to first order in $\bm v$. 

Deviations from the adiabatic approximation correspond to inelastic transitions between adiabatic states due to the variation of $\bm{R}_P(t)$. Let us evaluate the contribution of these inelastic transitions to (\ref{f:eq3}) when $\bm{R}_P(t)$ varies from 
$\bm{R}_0$ to $\bm{R}_1$. We study the evolution of a KS orbital equal to the adiabatic orbital $\varphi_{\bm{k}_0,i_0}$ for $\bm{R}_P(t_0)=\bm{R}_0$.  We set $\hat{\bm{v}}\cdot\bm{R}_P(t)=Z=v\,t$ and, to simplify our notations, hereafter only mention explicitly the dependence on $Z$.
For $Z>Z_0$, the KS orbital becomes $\psi_{\bm{k}_0,i_0}$, which we express as:
\begin{equation}
\psi_{\bm{k}_0,i_0}(Z) = \sum_{\bm{k}',j} c_{\bm{k}',j}(Z)\,\varphi_{\bm{k}',j}(Z)\, 
\exp\Big\{-{\im \over v} \int_{Z_0}^Z \dm Z' \varepsilon_{\bm{k}',j}(Z')  \Big\}
\label{noad:eq7}\end{equation}
with the initial condition $\psi_{\bm{k}_0,i_0}(Z_0)=\varphi_{\bm{k}_0,i_0}(Z_0)$. Using the time-dependent Schr{\"o}dinger equation, we obtain \cite{brajoa}:
\begin{equation}
{\dm \over \dm Z} c_{\bm{k}',j}(Z) =-\sum_{\bm{k},i} c_{\bm{k},i}(Z) \langle \varphi_{\bm{k}',j}(Z)| {\dm \over \dm Z} \varphi_{\bm{k},i}(Z) \rangle
\exp\Big\{-{\im \over v} \int_{Z_0}^Z \dm Z' \left[\varepsilon_{\bm{k},i}(Z') - \varepsilon_{\bm{k}',j}(Z')\right] \Big\}
\label{noad:eq8}\end{equation}
A first order approximation yields:
\begin{equation}
c_{\bm{k}',j}(Z_1) = \delta_{\bm{k}'j,\bm{k}_0i_0} - \int_{Z_0}^{Z_1} \dm Z \langle \varphi_{\bm{k}',j}(Z)| {\dm \over \dm Z} \varphi_{\bm{k}_0,i_0}(Z) \rangle
\exp\Big\{-{\im \over v} \int_{Z_0}^Z \dm Z' \left[\varepsilon_{\bm{k}_0,i_0}(Z') - \varepsilon_{\bm{k}',j}(Z')\right] \Big\}
\label{noad:eq9}\end{equation}
Now, the $v$ dependence of the excitations (i.e., the non adiabatic contributions) is entirely governed by the exponential terms in (\ref{noad:eq8}) or (\ref{noad:eq9}).
They oscillate rapidly for vanishing $v$, which quenches dramatically the transition probability when the two states are non degenerate. Then, as is well known, the transition probability does not vary as a power low in $v$ and is certainly not linear in $v$. This leads to a vanishing contribution to the Friction coefficient. However this is not the case for degenerate states or near-degenerate states: the transition probability is not quenched by the oscillations caused by the exponential if the energy difference between two states is of the order of $v$. We have to face this situation in our problem, since the occupied states belong to continua. We must, therefore, evaluate the corresponding contribution to the force (\ref{f:eq3}) at $Z_1$:
\begin{equation}
{\cal S}(Z_1) =\left.{\dm \over \dm Z}E(Z)\right|_{Z=Z_1} = - \int d\bm{r}\, \left[n_{\rm nonad}(\bm{r};Z_1) -
n_{\rm ad}(\bm{r};Z_1)\right]\, \left.{\dm V_{eP}(\bm{r};Z) \over \dm Z}\right|_{Z=Z_1} 
\label{noad:eq10} \end{equation}
where
\begin{eqnarray}
n_{\rm nonad}(\bm{r};Z_1) &=& \int_{FV} \dm \bm{k} \left| \psi_{\bm{k},i}(Z_1) \right|^2 \nonumber \\
n_{\rm ad}(\bm{r};Z_1) &=& \int_{FV} \dm \bm{k} \left| \varphi_{\bm{k},i}(Z_1) \right|^2
\label{noad:eq11} \end{eqnarray}
and the integration is inside the unshifted Fermi surface (FV).
Using (\ref{noad:eq9}) and keeping only the first order term:
\begin{eqnarray}
{\cal S}(Z_1) &=& \sum_j \int_{FV} \dm\bm{k} \int \dm\bm{k}' \, 2\Re\, \Big[ \langle \varphi_{\bm{k},i}(Z_1) |\left.{\dm V_{eP}(\bm{r};Z) \over \dm Z}\right|_{Z=Z_1} | \varphi_{\bm{k}',j} (Z_1)
\rangle \nonumber \\ 
& & \hspace{2cm} \int_{Z_0}^{Z_1} \dm Z \langle \varphi_{\bm{k}',j}(Z)| {\dm \over \dm Z} \varphi_{\bm{k},i}(Z) \rangle
\exp\Big\{-{\im \over v} \int_{Z_1}^Z \dm Z' \left[\varepsilon_{\bm{k},i} - \varepsilon_{\bm{k}',j} \right] \Big\} \, \Big]
\label{noad:eq12} \end{eqnarray}
Remembering that $\varphi_{\bm{k},i}$ is an eigenfunction of the static ($Z$ fixed) Schr{\"o}dinger equation, we obtain readily:
\begin{equation}
\langle \varphi_{\bm{k},i}(Z) |{\dm V_{eP}(\bm{r};Z) \over \dm Z}| \varphi_{\bm{k}',j} (Z)
\rangle = (\varepsilon_{\bm{k},i} - \varepsilon_{\bm{k}',j}) \langle \varphi_{\bm{k},i} (Z)|{\dm \over \dm Z} \varphi_{\bm{k}',j}(Z)
\rangle
\label{noad:eq13} \end{equation}
The matrix element in the r.h.s.\ depends only on properties of the adiabatic functions $\varphi_{\bm{k},i}$ and not on the velocity. However, from the preceding discussion, we know that inelastic contributions are only appreciable if $(\varepsilon_{\bm{k},i} - \varepsilon_{\bm{k}',j})$ is of the order of $v$. Furthermore, the integration over ${\bm{k}'}$ in (\ref{noad:eq12}) for a given ${\bm{k}}$
can be transformed into an integration over $\varepsilon_{\bm{k}',j}$. The transformation from one integration variable to the other involves only properties of the static electronic structure and is independent of the velocity. Again, the range of the integration over $\varepsilon_{\bm{k}',j}$ is of order $v$ around $\varepsilon_{\bm{k},i} = \varepsilon_{\bm{k}',j}$. As a consequence, the quantity ${\cal S}$ is of order $v^2$. 

Until now we have only considered the first order approximation to the $Z$-dependent problem as defined in (\ref{noad:eq9}). However, it can be easily verified that each increase in the perturbative order introduces an additional integration over the energy of intermediate states and, therefore, an additional factor of $v$. 
As a consequence, the non-adiabatic contributions to the force on the projectile are of order $v^2$ at least, which means that they do not contribute to the friction coefficient. 

We conclude that both $n_0$ and $n_{\bm v}$ (to first order in $v$) can
be determined exactly by the adiabatic SFS-KS scheme. So the latter
scheme provides an {\em exact} procedure to determine the friction coefficient.
\subsection{Corollary}
The SFS-KS scheme relies only on {\em local} properties of the
system, i.e., the determination of the density $n_{\bm v}$ only requires
information on the electronic state for a fixed value of $\bm{R}_P$. It does not
require information on the variation of any electronic quantity with $\bm{R}_P$.
Consequently, any alternative to (\ref{f:eq4})
for the evaluation of the friction coefficient must satisfy the same
condition. This provides a powerful tool to evaluate the validity of procedures
aiming at the evaluation of ${\cal F}_{\hat{\bm v}}$.

We may apply this condition to analyze the procedure of
Trail {\em et al.}\ \cite{trail1, trail2,trail3} (see also
Ref.~\onlinecite{luntz1}). The latter authors use the expression of
the friction coefficient proposed by Hellsing and Persson
\cite{hp84} (see also Ref.~\onlinecite{trail3}).
\begin{eqnarray}
{\cal F}_{\rm HP} &=& 2\pi\, k_F^2 \int \dm\hat{\bm{k}}_F
\int \dm\hat{\bm{k}}_F'\,
\left|\int \dm\bm{r} [\varphi^-_{\bm{k}_F'}(\bm{r})]^*\,
\varphi^+_{\bm{k}_F}(\bm{r})\right. \nonumber \\
& & \hspace{1.5cm} \left.\hat{\bm v}\cdot\bm{\nabla}_{\bm{R}_P}\right|_n
\, v_{KS}(\bm{r},\bm{R}_P;[n])\, \bigg|^2
\label{f:eq5}\end{eqnarray}
where $v_{KS}$ is the full KS potential, $k_F$ the Fermi momentum and
$\varphi^+$ (resp.\ $\varphi^-$) satisfies outgoing (resp.\ ingoing) boundary
conditions. From
the derivation of (\ref{f:eq5}) in Ref.~\onlinecite{hp84} (or the
alternative derivation in Ref.~\onlinecite{trail3}), it is not clear that it
consists in a first order approximation in $v$, though
when the impurity is moving through an homogeneous jellium, it can be proved
\cite{salin} that
(\ref{f:eq5}) is equivalent to (\ref{f:eq4}) to first order in $v$. A key
property, in the latter proof, is the invariance by translation of the free electron
gas state in the absence of an external potential. For the inhomogeneous case
such an equivalence has not been established. At any rate, in the latter case,
and in view of
the previous discussion, it is clear that the derivative of $v_{KS}$ with
respect to $\bm{R}_P$ in (\ref{f:eq5}) must be carried out for a constant
density, otherwise it would introduce a {\em non-local} contribution
(associated with the variation of $n$ when the projectile moves). However,
the authors of Ref.~\onlinecite{trail1, trail2,trail3, luntz1}  use a finite
difference method to calculate the derivative of $v_{KS}$: they determine the KS
potential for the projectile at rest and for two different values of $\bm{R}_P$,
say
$\bm{R}_P+\delta\bm{R}_P$ and $\bm{R}_P-\delta\bm{R}_P$. In so doing, they
include a term associated with the variation of the {\em function} $v_{KS}$ with
$\bm{R}_P$. The latter term, being non-local, introduces an error in the
evaluation of the friction coefficient.


\section{Illustration with a 1D model}

\begin{figure}
\centering
\scalebox{0.38}{\includegraphics{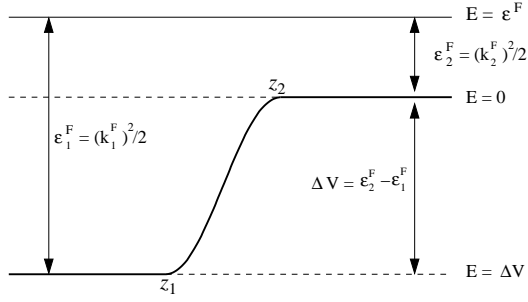}}
\caption{Sketch of the background potential used to create an inhomogeneous
jellium.}
\label{fig:fig1}
\end{figure}

\begin{figure}
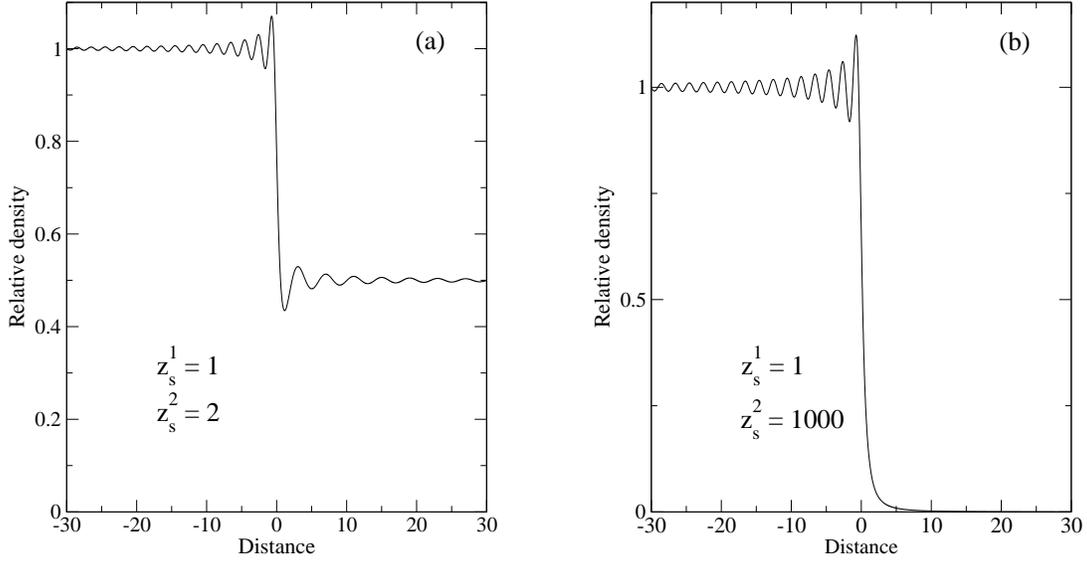

\vspace{1cm}
\centering
\scalebox{0.38}{\includegraphics{Fig2a.eps}}
\hspace{1cm}
\scalebox{0.38}{\includegraphics{Fig2b.eps}}
\caption{Density in the presence of the background potential (\ref{1D:eq2})
with $z_1=-0.5$, $z_2=0.5$, $z_s^1=1$ and (a) $z_s^2=2$ or (b) $z_s^2=1000$. The
density has been divided by the density of a uniform jellium with $z_s=1$.}
\label{fig:fig2}
\end{figure}

The previous discussion has raised two questions. Firstly, how important is
the error due to the non-local term in the evaluation of the friction
coefficient? Secondly, is expression (\ref{f:eq5}) valid for the inhomogeneous
case? To answer these questions, we have built a simple 1D model
amenable to an exact numerical solution.  Full details are given in the
Ancillary material\cite{salin}.
As we are concerned with general principles, our model needs not represent a
real system, although we keep speaking of ``electrons''. Though units are here
irrelevant, since we are only interested in relative results, we keep using
atomic units so that the ``electrons'' have a mass of 1. The interaction between
two particles located at $z$ and $z'$ is described by a screened regularized
coulomb potential:
\begin{equation}
 v_c^\pm(z,z')=\pm\, e^{-\alpha|z-z'|}/[|z-z'|+\epsilon]
\label{1D:eq1}\end{equation}
($v_c^+$ between electrons and $v_c^-$ between electrons and ``positive''
particles).
In the numerical application, we have used $\alpha=2$ and $\epsilon=0.1$. We
solve the problem at the level of the Hartree approximation.
The Fermi momentum for a 1D
paramagnetic homogeneous jellium with linear density $n$ is $k^F=n\pi/2$. We
introduce the Wigner-Seitz distance $z_s=1/n$, i.e., the length of the interval
enclosing one charge.
We create an
inhomogeneous jellium by introducing a background step potential defined by:
\begin{eqnarray}
v_{\rm step}(z) &=& 0.5\,\Delta V\, \{1+\cos[\pi(z-z_1)/(z_2-z_1)]\}
\qquad (z_1 \le z \le z_2) \nonumber \\
&=& \Delta V \qquad (z \le z_1) \nonumber \\
&=& 0 \hspace{0.4cm} \qquad (z \ge z_2)
\label{1D:eq2}\end{eqnarray}
We use throughout $z_1=-0.5$ and $z_2=0.5$.
The value of $\Delta V$ ($\Delta V<0$) is fixed by imposing values of $z_s^1$
and $z_s^2$ at $z\rightarrow-\infty$ and $z\rightarrow\infty$ respectively. The
Fermi energy, $\varepsilon_F$, is then given by
$\varepsilon^F=(k_1^F)^2/2+\Delta
V=(k_2^F)^2/2$, so that $n_1\ge n_2$ and $z_s^1\le z_s^2$. The energy diagram is
represented in
Fig.~\ref{fig:fig1}.
We introduce a background of positive particles such that
the Hartree potential is zero in the absence of impurity (i.e., the density of
positive particles is everywhere equal to that of the electrons). As
an
example, we plot in Fig.~\ref{fig:fig2} the density for $z_s^1=1, z_s^2=2$ and
$z_s^1=1, z_s^2=1000$. In the latter case, the behavior of the density
is qualitatively similar to that of a surface.

The interaction between an ``electron'' at $z$ and the impurity at $Z_P$ is
represented by the potential $v_c^-(z,Z_P)$ with $\alpha=4$ and $\epsilon=0.1$.
This potential supports one bound singly occupied state with energy -1.275.
For all results given below, the projectile is located at $Z_P=0$.
Calculation of the density in the presence of the impurity (even for
$v=0$) requires to confine the system into a box because the Hartree potential
behaves as $\sin(2k^F_{1,2}z + \gamma_{1,2})/z$ far from the impurity. We have
used a finite interval $z\in[-5,5]$. This does not invalidate our conclusions
since we are interested in the difference between two expressions of the
friction coefficient for a given system. We have checked that our conclusions
are independent of the box size and the implementation of the
contour conditions.

In a first step, we assume that the impurity, the step potential and the
associated background of positive charges move together against a uniform
jellium with $z_s=z_s^2$. The friction coefficient can then be calculated in
three different ways:

\begin{itemize}
\item[(i)] From the 1D expression equivalent to (\ref{f:eq4}):
\begin{eqnarray}
{\cal F}_{(i)} &=& \lim_{v\rightarrow 0} {1 \over v} \int_{-\infty}^{+\infty}
\dm z\, \left[n_v(z)-n_0(z)\right]
{\dm \over \dm Z_P} \Big[v_{eP}(z-Z_P) \nonumber \\
& & { } \hspace{6cm} +v_{\rm step}(z-Z_P)
+v_+(z-Z_P)
\Big]
\label{1D:eq3}\end{eqnarray}
where $v_{\rm step}$ and $v_+$ are respectively the step potential and potential
due to the background of positive charges.
\item[(ii)] It is easily shown\cite{salin} that it may be also calculated from
the energy
loss or gain associated with reflection and transmission by the potential. When
$v<k_2^F$:
\begin{equation}
{\cal F}_{(ii)} =
 {1 \over \pi} \left[(k_1^F)^2-(k_2^F)^2 + (k_1^F+k_2^F)^2\,P_r^F\right]
\label{1D:eq4}\end{equation}
where $P_r^F$ is the reflexion probability at the Fermi level for the static
case. This very simple expression gives us a good check on the accuracy of our
SFS calculations.
\item[(iii)] From the 1D version of (\ref{f:eq5}):
\begin{eqnarray}
{\cal F}_{(iii)} &=& 2\pi  \int \dm\bm{k} \int \dm\bm{k}'\,  \left|\int \dm z
\left[\varphi_{{\bm k}'}(z)\right]^*\, \varphi_{\bm k}(z) \right. \hspace{1cm}
\nonumber \\
& & \hspace{1cm}  \left. {\dm  \over\dm Z_P}\right|_n
v_{KS}(z-Z_P;[n(z,Z_P)]) \Big|^2\, \delta(\varepsilon_F-\varepsilon_k)\,
\delta(\varepsilon_F-\varepsilon_{k'})
\label{1D:eq5}\end{eqnarray}
As explained above, the derivative with respect to $Z_P$ is evaluated for a
constant density $n$, i.e., the quantities entering (\ref{1D:eq5}) must be
local.
\end{itemize}
The equivalence between (\ref{1D:eq5}) and (\ref{1D:eq3}) or (\ref{1D:eq4}) is
due to the fact that, under the present conditions, the transformation of the
system state from one position of the full external potential to another one
involves merely a translation (see Ref.~\onlinecite{salin} for more details).

\begin{table}
\caption{Friction coefficient when the impurity moves together with the
background potential. Results in the first
two columns are for double occupancy of the bound state and in the last two
columns for single occupancy. \label{tab1}}
\begin{center}
\begin{tabular}{|l|d|d|d|d|}
\hline
$z_s^1$ ; $z_s^2$&\multicolumn{1}{c|}{ 0.75 ; 1} &\multicolumn{1}{c|}{ 1 ; 1.5}
&\multicolumn{1}{c|}{ 1 ; 2} & \multicolumn{1}{c|}{1 ; 5} \\
\hline
${\cal F}_{(i)}$ (Eq.~\ref{1D:eq3}) & 0.576 & 0.472 & 0.545 & 1.06\\
${\cal F}_{(ii)}$ (Eq.~\ref{1D:eq4})& 0.573 & 0.466 & 0.546 & 1.07\\
${\cal F}_{(iii)}$ (Eq.~\ref{1D:eq5})  & 0.576 & 0.476 &  0.560 & 1.12\\ \hline
\end{tabular}
\end{center}
\end{table}
In Table~\ref{tab1}, we summarize the results obtained for some values of
($z_s^1$;$z_s^2$). Results are the same for the three expressions of the
friction coefficient within the accuracy of the calculations. This gives us
confidence in our numerical procedures.

We now consider the case of real interest for our model: that of the impurity
moving alone, against the step-potential and background of positive charges,
i.e., an impurity moving within an inhomogeneous jellium. The calculations only
differ from the previous ones by dropping $v_{\rm step}$ and $v_+$ in
(\ref{1D:eq3}) and noting that the dependence of $v_{KS}$ on $Z_P$ in (\ref{1D:eq5}), for a
constant density, arises entirely from $V_{eP}$. Also, (\ref{1D:eq4}) is no
longer valid. Results are given in Table~\ref{tab2}.

\begin{table}
\caption{Same as Table~\ref{tab1} when the impurity is moving with respect to
the fixed background potential. The last line gives the results
obtained by inclusion in (\ref{1D:eq5}) of the spurious non-local term when the
derivative of the Kohn-Sham potential is calculated along (\ref{1D:eq6}).
\label{tab2}}
\begin{center}
\begin{tabular}{|l|d|d|d|d|}
\hline
$z_s^1$ ; $z_s^2$ &\multicolumn{1}{c|}{ 0.75 ; 1} &\multicolumn{1}{c|}{ 1 ; 1.5}
&\multicolumn{1}{c|}{ 1 ; 2} & \multicolumn{1}{c|}{1 ; 5} \\
\hline
From Eq.~\ref{1D:eq3}  & 0.458 & 0.321 & 0.440 & 0.71  \\
\hline
From Eq.~\ref{1D:eq5}  & 1.944  & 2.36 & 2.59 & 4.41  \\
\hline
$\Delta v$ from (\ref{1D:eq6}) in (\ref{1D:eq5}) & 0.404 & 0.254 & 0.379 & 0.484
\\ \hline
\end{tabular}
\end{center}
\end{table}
The first observation is that (\ref{1D:eq5}) is no longer equivalent to
(\ref{1D:eq3}). Therefore, the expression of Hellsing and Persson does not
provide the correct first order in $v$ for the energy loss of an impurity in an
inhomogeneous medium. The error may be quite appreciable, as shown by our model.
Note that the trivial difference mentioned above between the expressions for a
moving and fixed background potential precludes a numerical error as the cause
of this discrepancy. The fact that our 1D model may be qualified as unrealistic
cannot infirm the relevance of our conclusion since a single counterexample is
sufficient to disprove an assertion.

We evaluate now the error incurred when taking into account the change in the
potential function in the expression of the potential derivative, as done
in the finite difference procedure of Trail {\em et al} \
\cite{trail1,trail2,trail3}. In place of the derivative of the KS potential for
a constant density in (\ref{1D:eq5}) we use the expression:
\begin{equation}
\Delta v = \left\{ v_{KS}(z, Z_P+h; [n_{Z_P+h}] )  - v_{KS}(z,Z_P-h; [n_{Z_P-h}] )\right\} /2h
\label{1D:eq6}\end{equation}
In actual calculations we have used $h=0.01$.
The two KS potentials in (\ref{1D:eq6}) are the result of a calculation where
the projectile is at $Z_P+h$ and $Z_P-h$ respectively.
For that reason, (\ref{1D:eq6}) includes a non-local contribution coming from
the modification of the potential function when moving the projectile from
$Z_P+h$ to $Z_P-h$. Results are given on the third line of
Table~\ref{tab2}. They differ strongly from the previous ones (second line in Table~\ref{tab2}), up to nearly
an order of magnitude. This difference is not
due to the simple algorithm
used in calculating the derivative as in (\ref{1D:eq6}). If we use the same
algorithm in expression (\ref{1D:eq5}), while keeping the density constant,
results agree with those given on the second line of Table~\ref{tab2}.
Changing $h$ (within reasonable bounds) does not change either the
conclusion. This demonstrates the significant error introduced by
the functional derivative of the Kohn-Sham potential with respect to density. Note that
the third line of Table~\ref{tab2} should not be compared with the first one
since it corresponds to an incorrect calculation of expression (\ref{1D:eq5}),
the latter being intrinsically incorrect for an inhomogeneous system, as
verified above. The two errors being totally unrelated, there is no reason why
they should compensate each other. The fact that the results of the
third line in Table~\ref{tab2} are closer to those of the first one must be
considered as merely accidental.
\section{Conclusion}

Starting from first principles, we have shown that the friction coefficient is,
in principle, exactly determined by an ensemble Kohn-Sham procedure, the
ensemble being defined by a Shifted Fermi Surface. We have shown that it
depends only on local properties of the impurity/target system in
the sense that its determination does not require information on the
variation of the system density with impurity position. It may be
of interest to check whether this conclusion could also be reached for other
applications in which the low energy behavior is determined through an
adiabatic approach.

To our knowledge, this important constraint has been overlooked. As a
consequence erroneous values of the friction coefficient have been obtained.
This is the case, for example, in the work of Trail
{\em et al.}\cite{trail1}. In addition, the latter authors use 
expression (\ref{f:eq5}) of Hellsing and Persson for the friction
coefficient, which, though correct for an impurity moving through an homogeneous
system, is not valid for the inhomogeneous case.

That the error may be dramatic is illustrated by the divergence found by Trail
{\em et al.}\ \cite{trail2,trail3}. They show, in their analysis, that the
divergence is caused by the derivative of the system spin with respect to
$\bm{R}_P$. The latter quantity is basically non-local (locality being used here
with the meaning defined above) and, therefore, the associated contribution to
friction is spurious. The latter authors wrongly attribute the divergence to a
breakdown of the adiabatic approximation. However, if the evolution for
$v\rightarrow 0$ is not adiabatic, this means that there is a discontinuity in
the evolution of the system. In the case studied by Trail {\em et al}, no such
discontinuity exists. So, the divergence can only be the result of an error in
the evaluation of the friction coefficient, as confirmed by our analysis.

Another conclusion can be drawn with respect to the evaluation by Luntz  {\em et
al.}~\cite{luntz2} of the ``local approximation for friction''
(LDAF)\cite{juar08}. In the latter approximation, the friction coefficient is
evaluated as a weighted average of the friction coefficient for the impurity in
a homogeneous electron gas having the local density at each point of the
inhomogeneous system. The calculations of Luntz  {\em et al.}~\cite{luntz2},
based on (\ref{f:eq5}) and (\ref{1D:eq6}), are incorrect for the inhomogeneous
case and, therefore, the difference they find with respect to the LDAF cannot be
considered as an evaluation of the error incurred when using the LDAF.

Finally, we may remark that the evaluation of friction with present day band
structure codes requires an SFS calculation since, up to now, no well founded
alternative exists.

\begin{acknowledgments}
The author would like to thank M.\ Alducin, H.F.\ Busnengo, R.\ D\'\i ez-Mui\~
no, and J.I.\ Juaristi for stimulating this work and for
their useful comments. Thanks are due to V.H.~Ponce for a critical reading of
the manuscript.
\end{acknowledgments}

\end{document}